\documentclass{aa}
\usepackage{natbib}
\usepackage{psfig,graphicx,epsfig}

\def\gsim{\;\lower4pt\hbox{${\buildrel\displaystyle >\over\sim}$}\,}
\def\lsim{\;\lower4pt\hbox{${\buildrel\displaystyle <\over\sim}$}\,}

\def \xmm {{\em XMM-Newton}}

\def \hcm {\hbox {\ifmmode $ atom cm$^{-2}\else atom cm$^{-2}$\fi}}

\def \arcsec {\hbox{$^{\prime\prime}$}}

\def\approxgt{\mathrel{\hbox{\rlap{\lower.55ex \hbox {$\sim$}}
        \kern-.3em \raise.4ex \hbox{$>$}}}}
\def\approxlt{\mathrel{\hbox{\rlap{\lower.55ex \hbox {$\sim$}}
        \kern-.3em \raise.4ex \hbox{$<$}}}}

\newcommand\phib{\phi_{\bf B}}
\newcommand\phigb{\phi_{\nabla |{\bf B}|}}
\newcommand\gb{\nabla |{\bf B}|}

\def\lsim{\;\raise0.3ex\hbox{$<$\kern-0.75em\raise-1.1ex\hbox{$\sim$}}\;}
\def\gsim{\;\raise0.3ex\hbox{$>$\kern-0.75em\raise-1.1ex\hbox{$\sim$}}\;}

\def\beq{\begin{equation}}
\def\enq{\end{equation}}
\def\begar{\begin{eqnarray}}
\def\endar{\end{eqnarray}}
\def\mathnew{\mathsurround=0pt}
\def\simov#1#2{\lower .5pt\vbox{\baselineskip0pt \lineskip-.5pt
        \ialign{$\mathnew#1\hfil##\hfil$\crcr#2\crcr\sim\crcr}}}

\def \xmm {{\it XMM-Newton}}

\begin{document}


\title{Constraints on local interstellar magnetic field from non-thermal emission of SN1006}


\author{F. Bocchino\inst{1}
\and S. Orlando\inst{1}
\and M. Miceli\inst{2}
\and O. Petruk\inst{3}
}

\institute{
       INAF-Osservatorio Astronomico di Palermo, Piazza del Parlamento 1,
       90134 Palermo, Italy
\and 
       Dipartimento di Scienze Fisiche ed Astronomiche, 
       Sezione di Astronomia, Universit\`a  di Palermo, 
       Piazza del Parlamento 1, 90134 Palermo, Italy
\and
       Institute for Applied Problems in Mechanics and Mathematics, 
       Naukova St. 3-b, 79060 Lviv, Ukraine
}

\date{Received  / Accepted }

\authorrunning{Bocchino et al.}
\titlerunning{Local {\bf B} field around SN1006}

\abstract
  {The synchrotron radio morphology of bilateral supernova remnants depends on the mechanisms of particle acceleration and on the viewing geometry. However, unlike X-ray and $\gamma$-ray morphologies, the radio emission does not depend on the cut-off region of the parent electron population, making it a simpler and more straightforward tool to investigate the physics of cosmic ray production in SNRs.}
  {Our aim is to derive from the radio morphology tight constraints on the direction of the local magnetic field and its gradient, and on the obliquity dependence of the electron injection efficiency. }
  {We perform a set of 3D MHD simulations describing the expansion of a spherical SNR through a magnetized medium with a non-uniform ISMF. From the simulations, we derive non-thermal radio maps and compare them with observations of the SN1006 remnant.}
  {We find that the radio morphology of SN1006 at 1 GHz is best-fitted by a model with quasi-parallel injection efficiency, a magnetic field aspect angle of $38^\circ\pm 4^\circ$ with the line of sight, and a gradient of the field strength toward the galactic plane, higher then the expected variations of the large scale field of the Galaxy.}
  {{  We conclude that the radio limbs of SN1006 are polar caps, not lying in the plane of sky}. The study of the synchrotron radio emission of SNRs is of crucial importance to derive information on the galactic magnetic field in the vicinity of the remnants, and more hints on the actual injection efficiency scenario.}


\maketitle

\section{Introduction}

The radio morphology of supernova remnants (SNRs) may be very informative
on the conditions of the magnetized environments in which the blast-wave
expands
and, more in particular, on the acceleration processes
which occurs at the shock front, and which give rise to the energetic
electrons ultimately responsible of the synchrotron emission in the
radio and X-ray band. In particular, the radio emission, unlike the X-ray emission which depends  on both the
efficiency of electron injection into the acceleration process { (i.e. the fraction of particles injected from the thermal gas, as in e.g. \citealt{bgv05})} and on
the rapidity of acceleration to high energies, is insensitive to acceleration-rate issues. In fact, the time required to accelerate an electron
emitting synchrotron emission at a frequency $\nu$ is about $3 \times 10^{-3} \eta (\nu/GHz)^{0.5} (B/100 \mu G)^{-1.5}(V_s/3,000$ km s$^{-1})^{-2}$ yr (\citealt{uat07}), where B is the magnetic field, $V_s$ is the shock speed, $\eta$ is { the ratio between the mean free path of the particles along the magnetic field line to the gyroradius, and the parameters are normalized to typical young SNR values. For the archetypical remnant SN1006, we have $B \sim 100$ $\mu$G (\citealt{vbk05}, \citealt{b06}, \citealt{mab10}, \citealt{pbb11}), $V_s \sim 4000$ km s$^{-1}$ (\citealt{mgr93}, \citealt{kpl09}), and $\eta \sim 1-10$ (\citealt{pbb11}), thus, for radio emitting electrons, the resulting acceleration time (less than a year) is
extremely rapid compared to SNR evolutionary timescale ($\sim 1$ kyr).} As a result, variations of radio morphology with obliquity point to electron injection physics
alone, making it the best regime by which to investigate this issue.

In this context, the remnants showing a radio shell
with two opposite bright and regular limbs separated by two minima of
emission (often named bilateral or barrel-shaped or bipolar, BSNRs hereafter,
\citealt{kc87}, \citealt{gae98}) are considered ideal laboratories,
because their morphology is definitely not heavily affected by small
scale inhomogeneities which may make the interpretations rather difficult.  A point-like
supernova explosion which occurs in a uniform magnetized medium with constant value
and direction of the interstellar magnetic field (hereafter ISMF) yields a symmetric radio BSNR whose bright
limbs are located where the magnetic field is parallel or perpendicular
to the shock normal, if the injection efficiency is quasi-parallel or
quasi-perpendicular/isotropic\footnote{{Here, perpendicular and parallel refers to the angle between the { shock normal} and the pre-shock magnetic field (this
angle is called the obliquity angle). 
In particular, the injection is called quasi-parallel (quasi-perpendicular) if its efficiency is maximum where the obliquity angle is $0^\circ$ ($90^\circ$, \citealt{fr90}). The injection is called isotropic if its efficiency does not depend on the obliquity angle  (\citealt{ebj95}, \citealt{vbk03}).}}, respectively, and if the ISMF direction
lies in the plane of sky, whereas different configurations occurs at
different aspect angles (\citealt{fr90}; \citealt{obr07})

However, in the real universe, many SNRs evolve in a non-uniform ISM (see e.g. \citealt{hp99} for an analytical treatment of SNR in nonuniform medium) and BSNRs are therefore often asymmetric.
\citet{obr07} (hereafter Paper I) have generalized the study
of \citet{fr90} to these cases by considering the explosions occurring in a large
scale gradient either of density or magnetic field, showing that most of the asymmetries observed in true BSNRs can be recovered with this model. In particular, the radio
morphology loses one axis of symmetry, and the limbs are not equally
bright (if the gradient runs across the limbs) or they are not opposite
and converge on the side in which either the density or the magnetic field
is increasing (if the gradient runs between the limbs).

It is clear that the morphology of BSNRs is tightly coupled with
the magnetized environments in which the shock expands, and it is
of particular interest to note here the preference of BSNR symmetry
axis\footnote{The axis running parallel to the limbs in the plane of the sky.} to be oriented parallel to the galactic plane, as reported by
\citet{gae98}. It seems therefore possible to study their morphology
to derive the geometry of the surrounding magnetic field, and thus shedding
more light on the mycrophysics of the particle acceleration processes
which undergoes at the shock front. In this respect, BSNRs may be considered as a probe for structure of Galactic MF on scales of $\sim 10$ pc, much lower than the expected large scale variation of the field predicted by global models (\citealt{ps03}, \citealt{kst07}, \citealt{srw08}).

The remnant of SN1006 seems to be the object in which this kind of study
may be more fruitful. The uniform environment and the bright limbs visible
in most of the electromagnetic spectrum makes it a real case study in
the field of particle acceleration mechanisms in strong shocks.  Indeed,
\citet{rbd04}, using a simple and powerful geometrical argument applied to
the \xmm\  X-ray image of SN1006, based on the ratio between the central
and the rim luminosity, showed that the bright limbs are polar caps
(instead  of an equatorial belt) and that, therefore, the magnetic field
is oriented across the limbs, in the NE-SW direction. This argument
seems to break the dicotomy between the two competing scenarios of the
dependence of the injection efficiency on the obliquity angle $\theta$
(the angle between the magnetic field direction and the shock normal),
preferring a situation in which the injection is most efficient when
the field is along the shock normal (quasi-parallel scenario) over the
situation in which the field is perpendicular (quasi-perpendicular
scenario, see \citealt{fr90}).

On one hand, these findings were somehow expected since \citet{vbk03}
already pointed out that in SN1006 the injection should be maximum at
parallel shocks. However, in the light of the uncertainties related to the
details of the acceleration processes, several authors still considered
the quasi-perpendicular scenario a viable option: \citet{fr90} argued
against the quasi-parallel scenario pointing out that quasi-parallel
models often give rise to unobserved morphologies in the radio band (see also \citealt{obr07});
\citet{yyt04} still considered both models to explain the observed
width of selected filaments of SN1006 observed by \citet{byu03},
whereas \citet{ah07} developed a quasi-perpendicular model which agrees
very well with the same data.



Recently, \citet{pdc09} have further investigated this issue, showing
that, in the framework of a simple model of SN1006 in terms of a
point-like explosion occurring in a uniform density and uniform magnetic
field medium, there is no way to reconcile the quasi-parallel scenario and
the SN1006 morphology as observed in the radio band. { The same conclusion was reached more recently by \citet{svr10}}. The contradiction
in the interpretation of the radio morphology (suggesting isotropic or quasi-perpendicular injection
efficiency scenario) and X-ray morphology (suggesting quasi-parallel
scenario), and the fact that SN1006 shows both the kinds of asymmetries discussed in Paper I (converging limbs and different surface brightness), has prompted us to investigate the effects of non-uniformity of
the environment on the observed properties of this remnant, capitalizing
on the work of \citet{obr07} on asymmetric BSNRs. 


In Sect. 2, we briefly describe the MHD model used to
reproduce the remnant of SN1006, which includes a small gradient of the
magnetic field. In Sect. 3,
we introduce the methodology for the comparison between modeled and
observed radio images of SN1006, while in Sect. 4 we
discuss the results of the comparison, showing that, not only the
model with the gradient of $|{\bf B}|$ is able to reconcile, for the
first time, the interpretation of radio and X-ray morphology of the remnant, but it also
provides stringent constraints on the overall geometry of the field in
the vicinity of the remnant.

\section{The model}

\begin{table}
\caption{MHD models of SN1006 used in this work}
\label{models}
\medskip
\centering\begin{minipage}{8.7cm}
\begin{tabular}{lcccc} \hline
  Name & $\nabla |{\bf B}|$\footnote{The 
  relative variation of the magnetic field strength over a scale of 10 pc} \\ \hline

UNIFORM\_B &  1.0 \\
GRAD1      &  1.2 \\ 
GRAD2      &  1.4 \\ 
GRAD3      &  2.0 \\ 
GRAD4      &  4.7 \\ 
\hline
\end{tabular}
\end{minipage}
\end{table}

We adopted the 3D MHD SNR model described by \citet{obr07} to which the reader is referred for further details, and we use the FLASH code (\citealt{for00}) for the model implementation.
Since there is accumulated evidence that the density around SN1006 is
fairly constant, with the exception of the NW sector where an encounter
with a dense cloud is occurring (\citealt{abd07}, \citealt{mbi09}), we
argue that a model with a $|{\bf B}|$ gradient is more appropriate than that with a density gradient to describe
the asymmetries of the limbs of SN1006. Given the pronounced slantness of the limbs, we conclude, following Paper I, that a prominent component of $\gb$ is directed parallel to the limbs (and therefore perpendicular to the galactic plane). 
The remnant was modeled as a
point-like explosion of 1.4 solar masses of ejecta having a kinetic
energy of $E=1.3\times 10^{51}$ erg occurring in a uniform density
medium with $n_0 = 0.05$ cm$^{-3}$ (\citealt{abd07}). The ISMF has a value of 30 $\mu$G in
the environment of the explosion site. Such a high value for the location of SN1006 in the Galaxy, has been chosen in order to take into account, as a first approximation, the effects of magnetic field amplification, so to have a post-shock $|{\bf B}|$ of the order of a few hundreds of $\mu$G in our simulations, in agreement 
with observations of some SN1006 filaments (e.g. \citealt{yyt04}).
The direction of MF is along the X axis while its gradient is along the Z (X) axis in case of quasi-parallel (quasi-perpendicular) injection efficiency (see. Fig. 1 of Paper I for more details). 
$\nabla |{\bf B}|$ values used in our simulations are reported in
Table \ref{models}. The simulations were
stopped at $t=1000$ yr, checking that the shock velocity and remnant
radius are compatible with the observed values (4600 km s$^{-1}$ and
9.6 pc, respectively).

{  We synthesized the radio emission using the REMLIGHT code of \citet{opb11}, to which the reader is referred for more details. In particular, the synchrotron radiation from relativistic electrons distributed with
a power law spectrum $N(E) \propto K E^{-s}$ is computed according to \citet{gs65}
with the formula $i(\nu) \propto K B_{\perp}^{(\alpha+1)}
\nu^{-\alpha}$, where $\alpha = (s-1) / 2$, and $B_{\perp}$ is the component of the magnetic
field perpendicular to the line-of-sight (LoS). The coefficient $K \propto \epsilon V_{sh}^{-b}$
includes all the dependencies from the obliquity and the shock velocity ($V_{sh}$) as described by \citet{obr07}, and, in particular, the injection efficiency is $\epsilon \propto sin^2 \Theta_{Bn2}$ and $\epsilon \propto cos^2 \Theta_{Bn2}$ in the quasi-perpendicular and quasi-parallel case, where $\Theta_{Bn2}$ is the obliquity angle between the shock normal and the post-shock magnetic field. { This particular choice of the obliquity dependance derives from the work of \citet{fr90}, and it is largely adopted in the literature (e.g. \citealt{obr07}, \citealt{svr10}, \citealt{opb11}; for alternative formulations see \citealt{vbk03} and \citealt{pbb09})}.
The parameter $b$ has been chosen equal to -1.5, reflecting a situation in which stronger shocks inject particle more efficiently, but we have verified that our results are not affected by choices of $b$ in the range (-1.5, +2.0), the range explored by \citet{obr07}}.


\section{Comparison between models and observations}

\subsection{The angles defining the viewing geometry}

\begin{figure}
  \centerline{\psfig{file=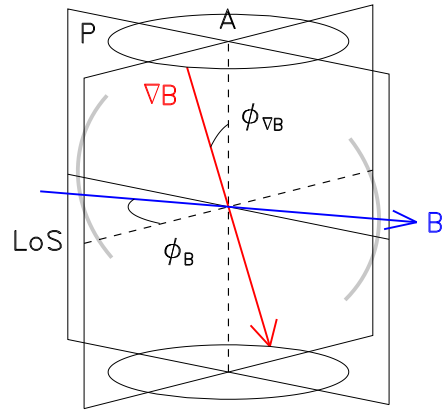,width=9.0cm}}
  \caption{The angles defining the viewing geometry, namely $\phi_{\bf B}$ (the magnetic field aspect angle) and $\phi_{\nabla |{\bf B}|}$ (the rotation angle of the gradient). $P$ is the plane of the sky, on which the SN1006 are projected (sketched as light grey). See text for more details.}

  \label{angles}
\end{figure}

Several works in the literature show that the observed morphology
of a SNR emitting synchrotron emission is strongly affected by the
angle between the direction of {\bf B} and the line
of sight, $\phi_{\bf B}$. Moreover, in our circumstance, since we
have also a gradient of the magnetic field, we are forced to consider
the dependence of the observed morphology from the angle between the
direction of the gradient and the plane of the sky, $\phi_{\nabla |{\bf
B}|}$. The definition of $\phi_{\bf B}$ and $\phi_{\nabla |{\bf B}|}$
are sketched in Fig. \ref{angles}. To the purpose of this work,
we have chosen to synthesize our maps in the following way: the starting configuration has the MF perpendicular to the LoS ($\phi_{\bf B}=90^\circ$), and $\nabla |{\bf B}|$ lying in the plane of the sky (the plane $P$ in Fig. \ref{angles}) and parallel to the lobes. First,
we apply a rotation of $\phi_{\nabla |{\bf B}|}$ degrees around the axis
in the plane of the sky passing through the center of the remnant and perpendicular to the limbs, where positive angles means that region of increasing {\bf B} are closer
to us. Then, we apply a rotation of $90^\circ - \phi_{\bf B}$ degrees around the
axis in the plane of the
sky passing through the remnant center, and parallel to the limbs (axis A in Fig. \ref{angles}). Others rotation schemes gives similar results. The adopted values
of $\phi_{\bf B}$ are from $0^\circ$ to $90^\circ$ in step of $2^\circ$
and the ones of $\phi_{\nabla |{\bf B}|}$ are  from $0^\circ$ to $90^\circ$
in step of $15^\circ$. Therefore, for each model, we have generated 315
maps encompassing all the combinations of the relevant angles.

\subsection{The morphological parameters}

\begin{figure}
  \centerline{\psfig{file=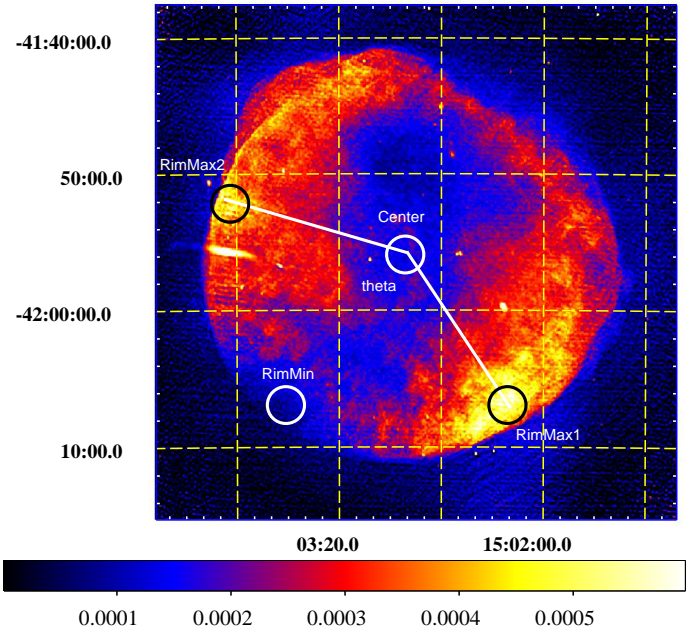,width=9.0cm}}
  \caption{A radio map of SN1006 at 1.4 GHz, with the four region in which the
  surface brightness of the remnant has been averaged for the purpose
  of computation of the four morphological parameters: $A=RIM_{max}/RIM_{min}$,
  $C=center/RIM_{max}$, $R_{max}=RIM_{max1}/RIM_{max2}$ and $\theta_D$, the aperture angle between the limb maxima. Adapted from \protect\citet{pdc09}.}

  \label{radiomap}
\end{figure}

The exploration of the parameter space is a challenging task, because
it involves the morphological comparison between many model maps and the
true SN1006 image. We devised a simple procedure which involves the
comparison of the value of 4 morphological parameters calculated both
in the synthesized radio emission maps and on the observed radio map
of SN1006. The parameters are: the ratio between the maximum and the
minimum of surface brightness ($S_b$) around the rim ($A$); the ratio
between the maximum around the rim and the center of the remnant ($C$),
the ratio of $S_b$ of the two bright limbs ($R_{max}$); the angular
separation between the limbs maxima ($\theta_D$). The parameters $A$, $R_{max}$
and $\theta_D$ have been also used in \citet{obr07}, and we refer to
that paper for further discussion. The $C$ parameter is introduced here
to measure the luminosity contrast between the brightest rim and the
center. Note that the $C$ parameter does not correspond to the $R_{\pi
/3}$ parameter used by \citet{rbd04} to exclude the equatorial belt
scenario for SN1006, because $C$ is measured in a small circular region
in the radio map (see below). While $R_{max}$ and $\theta_D$ gives a measure of the asymmetries (if any) in the remnant morphology, $A$ and $C$ give the idea of the remnant being centrally filled ($C>1$, $A\sim 1$) or barrel-shaped ($C<1$, $A > 1$). { This choice of evaluating the emission in selected regions makes the method less sensitive to the exact functional form of the obliquity dependance of the injection efficiency introduced in Sect. 2.}

The parameters $A$, $C$ and $R_{max}$ are
measured for SN1006 using the radio map of \citet{pdc09}, by averaging the $S_b$
value on circular regions of 45\arcsec radius (Fig. \ref{radiomap}). The
values we got are $A=2.7\pm0.2$, $C=0.36\pm0.03$, $R_{max}=1.2\pm0.1$,
and $\theta_D=135^\circ\pm10^\circ$. For the measurement of the parameters
in the synthesized radio maps we used  an automated procedure to
find the maximum of the two limbs, the minimum between limbs along the
rim and the central position. Then we used an average in circular regions
whose radius is the same percentage of the remnant radius used in the true
radio image (5\%) to minimize fluctuations. Since the model images have been synthesized using
the same number of pixels per remnant radius as the true  radio map, this procedure ensure
that the model and observed values of the parameters are comparable.

\subsection{Test of the procedure}

\begin{figure}
  \centerline{\psfig{file=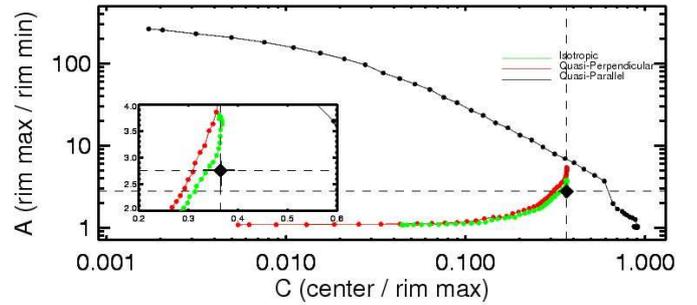,width=9.0cm}}
  \caption{Comparison between synthesized radio maps in the case of a uniform {\bf B} at an age of 1000yr (black, green and red dots for quasi-parallel, quasi-perpendicular and isotropic injection efficiency, respectively) and the observed radio map (black diamond, the inset show a zoom of the plot around this point) by means of the use of the A and C morphological parameters introduced in Fig. \protect\ref{radiomap}. Each curve represent the values spanned by the parameters when the aspect angle $\phi_{\bf B}$ goes from $0^\circ$ (rightmost part) to $90^\circ$ (leftmost part).}

  \label{nograd}
\end{figure}

To test the procedure, we run the model with the initial parameters discussed in Sect. 2 but assuming a uniform magnetic field, in the quasi-parallel, quasi-perpendicular and isotropic injection efficiency scenarios. This situation is already somehow in contradiction with the observed values $\theta_D=135^\circ$ and $R_{max}=1.2$, because $\gb = 0$ implies $\theta_D=180^\circ$ and $R_{max}=1$. However, this MF configuration has been used in the literature, and it is a good test case for our procedure.
In Fig. \ref{nograd}, we show the $A-C$ scatter plot for various $\phi_{\bf B}$ angles,
in the uniform {\bf B} case and quasi-parallel, quasi-perpendicular and
isotropic scenarios\footnote{In the uniform {\bf B} case, $R_{max}=1$
and $\theta_D=180^\circ$ by definition.}. In this case, the plot suggests that
quasi-parallel models does not reproduce at all the observed parameters,
unlike the quasi-perpendicular and isotropic scenarios, with a best-fit $\phi_{\bf
B} \approx 70^\circ$. This is in excellent agreement with the work of \citet{pdc09}, which derived $\phi_{\bf
B}$ by comparing the azimuthal brightness profile with models in uniform magnetic field. Therefore, we conclude that our method is robust.

\subsection{Results}

\subsubsection{Polar caps}

\begin{figure}
  \centerline{\psfig{file=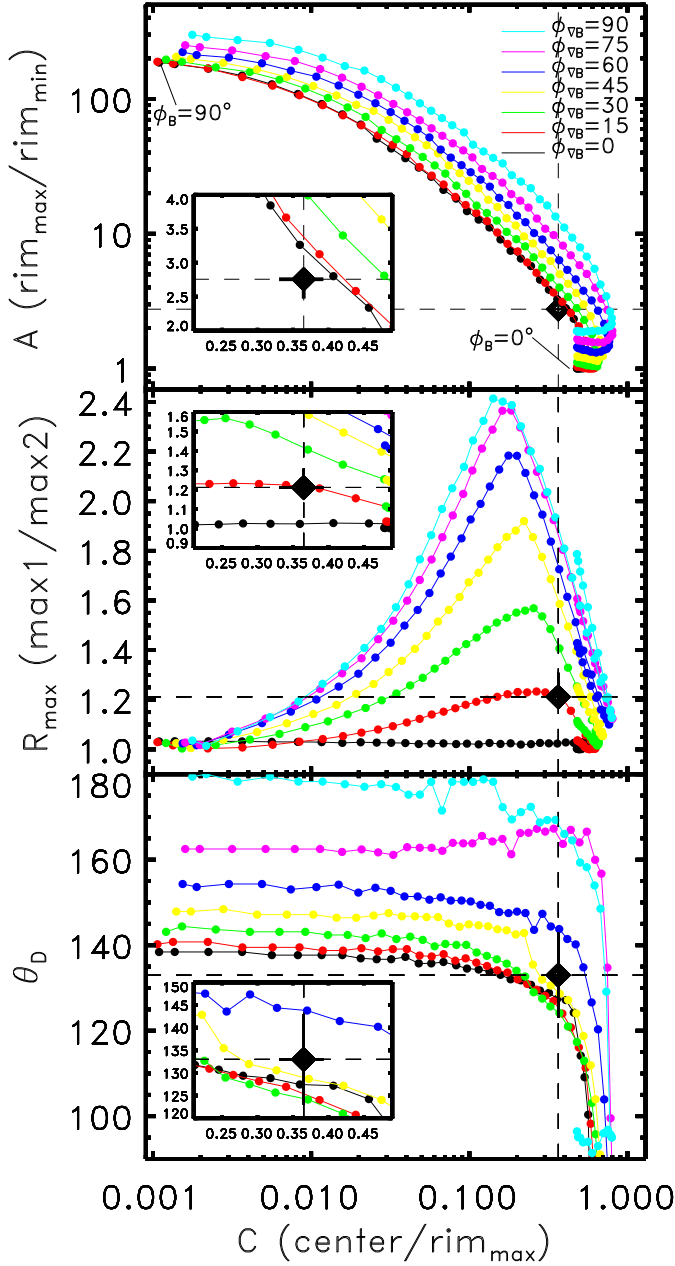,width=9.0cm}}
  \caption{{\em Top panel:} $A-C$ scatter plot for our model GRAD2 (Table \protect\ref{models}), assuming
  quasi-parallel injection and an age of 1000 yr. Different colors
  correspond to different values of $\phi_{\nabla
|{\bf B}|}$ (from $0^\circ$ to $90^\circ$ in steps of $15^\circ$), whereas each dot corresponds to a given value of $\phi_{\bf
B}$ (from $0^\circ$, right, to $90^\circ$, left, in steps of $2^\circ$). We overplotted
the (A,C) values measured for SN1006 (diamonds; the inset shows a zoom near this
point). 
{\em Middle panel:} Same as top panel but for the $R_{max}-C$ pair.
{\em Lower panel:} Same as top panel but for the $\theta_D-C$ pair.}

  \label{resqpar}
\end{figure}

In Fig. \ref{resqpar}, we show the $A-C$, $R_{max}-C$ and $\theta_D-C$
scatter plots computed in the synthesized radio maps of our SN1006 model
including a weak gradient of the magnetic field (GRAD2 in Table \ref{models}, i.e. {\bf B} varies of a
factor 1.4 over 10 pc scale), {  and assuming that the electron injection efficiency is maximum where the obliquity angle is $0^\circ$ (quasi-parallel scenario). This configuration has also been referred to as ``polar caps".} We
also overplotted the values of the parameters derived using the true
radio map of the remnant of \citet{pdc09}. Though the agreement is not
exactly perfect in all the plots, we note that we can define
a very limited region of the parameter space ($\phi_{\bf B}, \phi_{\nabla
|{\bf B}|}$) which is compatible with the observed values of $A$, $C$,
$R_{max}$ and $\theta_D$. This means that the observed radio morphology
of SN1006 is overall compatible with a quasi-parallel scenario for this
remnant, if we include a weak gradient of {\bf B}.

A remarkable result is
that the comparison between the quasi-parallel model and the observation
strongly exclude a situation in which the polar caps are in the plane of
the sky ($\phi_{\bf B}=90^\circ$) or along the line of sight ($\phi_{\bf
B}=0^\circ$). The latter geometry would cause a centrally brightened radio morphology
instead of two limbs, as already discussed by \citet{fr90} and \citet{obr07}. The
best-fit values of the relevant angles derived from Fig. \ref{resqpar},
and a conservative estimate of their uncertainties, are $\phi_{\bf
B}=38^\circ\pm 4^\circ$ and $\phi_{\nabla |{\bf B}|}=15^\circ\pm 15^\circ$.

\begin{figure}
  \centerline{\psfig{file=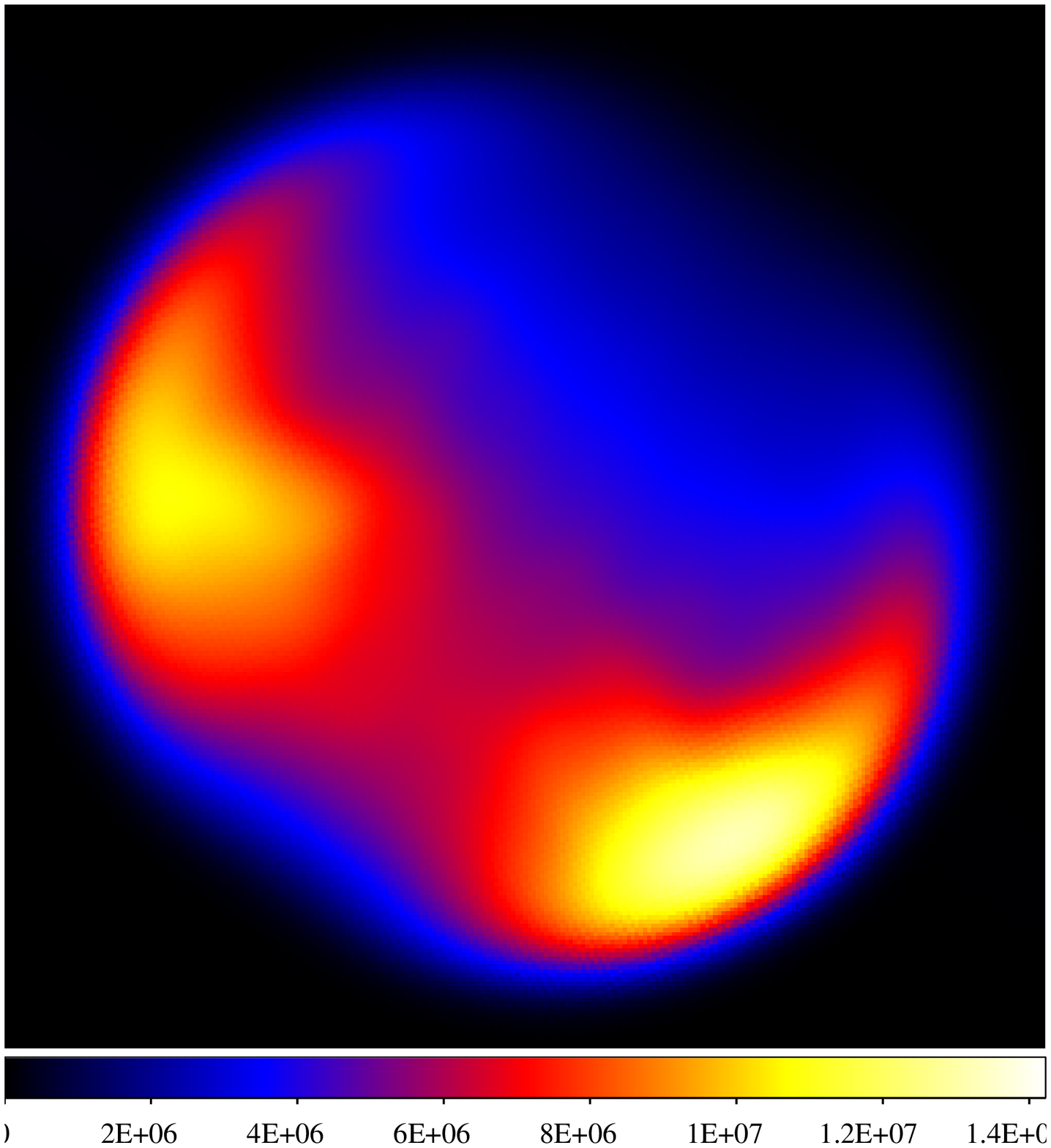,width=7.0cm}}
  \centerline{\psfig{file=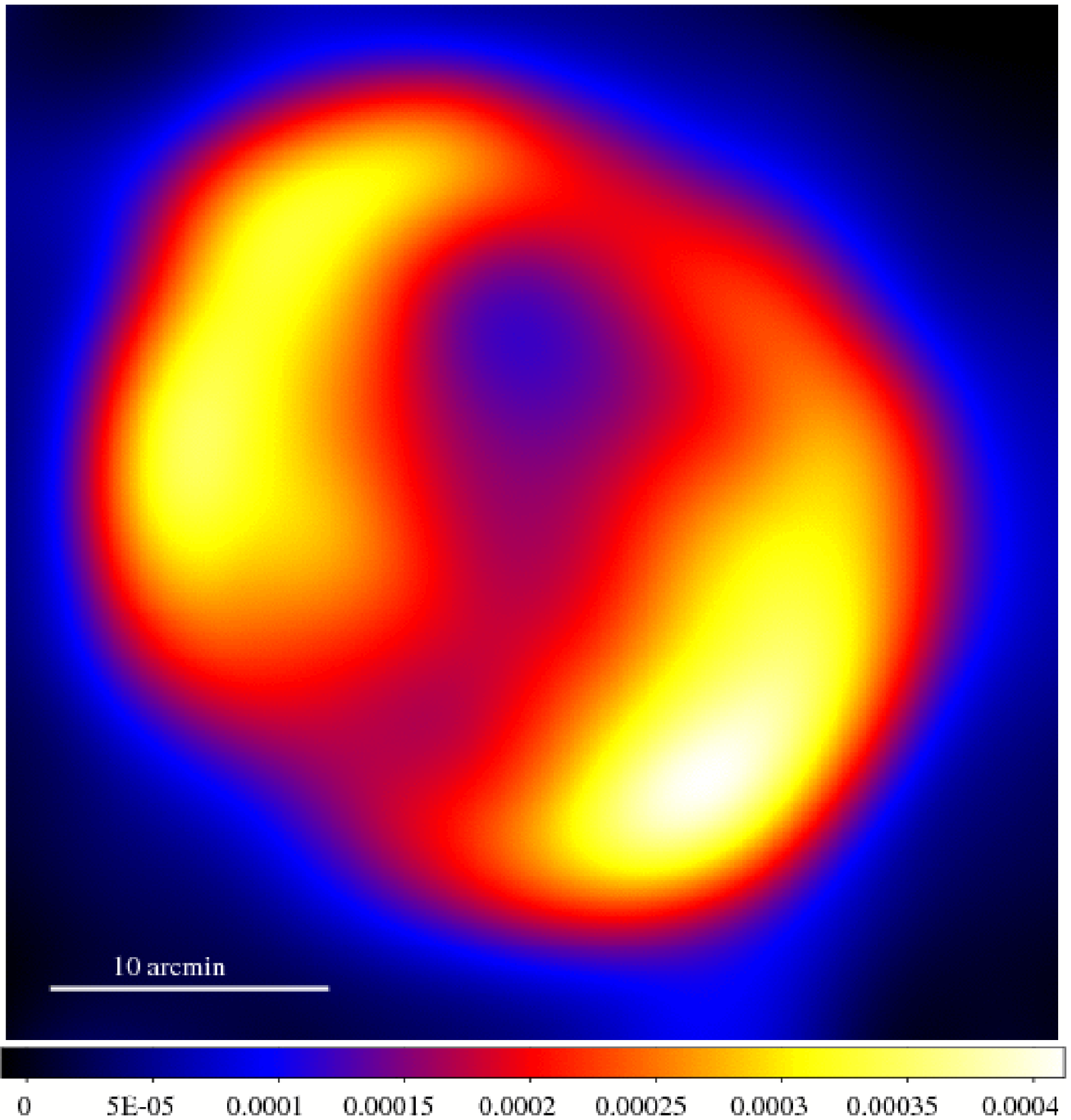,width=7.0cm}}

  \caption{{\em Top panel:} Synthesized radio image at 1 GHz for our best fit model (GRAD2 in Table \protect\ref{models}, and quasi-parallel injection efficiency scenario, $\phi_{\bf
B}=38^\circ$ and $\phi_{\nabla |{\bf B}|}=15^\circ$), smoothed with a sigma of $2^\prime$. {\em Bottom panel:} radio map of SN1006 of Fig. \ref{radiomap}, but smoothed with the same sigma used for the model image.}

  \label{obsmod}
\end{figure}

The synthesized radio map of the best fit model is shown in
Fig. \ref{obsmod} (top panel), along with the observed radio map (bottom panel). {  Both maps have been slightly  
smoothed ($2'$ sigma, $\sim 1/15$ the remnant size) to focus the comparison on
the general obliquity trends. In fact, we are not interested in reproducing the fine structure of the thermal emission from the ejecta which is visible in the high-resolution radio observation. We note that} the large scale structures of the observed radio emission
are very well recovered by the best-fit model and the two images looks
similar.

The comparison with other gradients, namely the GRAD1 (weaker gradient), GRAD3 and GRAD4
models (stronger gradient), shows that it is not possible to find a satisfactory fit for
all the parameters. In particular, for higher values of the gradient,
the angle $\theta_D$ is always smaller than observed in most of the cases, and for lower values is always larger, so GRAD2 is the only model which
gives us an overall good fit. The variation of the MF in SN1006 can therefore be bracketed between 1.2 and 2.0 (see Table \ref{models}).

\subsubsection{Equatorial belt}

\begin{figure}
  \centerline{\psfig{file=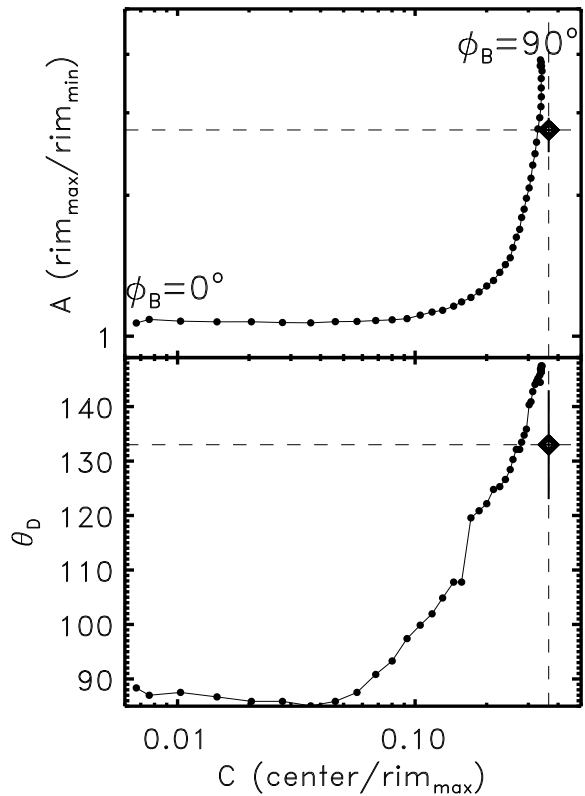,width=8.0cm}}
  \caption{{\em Top panel:} Same as Fig. \protect\ref{resqpar} (top panel)
  but for the GRAD2 model and quasi-perpendicular scenario. There is no $\phigb$ angle shown in
  this case, because {\bf B} is always aligned with $\gb$. {\em Lower
  panel:} Scatter plot of $\theta_D-C$ parameters. We overplotted the
  values observed in SN1006 (diamonds).}

  \label{resqper}
\end{figure}

{  In this section, we explore the same MHD model as in the previous section, but now assuming that the electron injection efficiency is maximum where the obliquity angle is $90^\circ$ (quasi-perpendicular scenario). This configuration has also been refereed to as ``equatorial belt".
We produced the synthetic images and we repeated the analysis described in the previous
paragraphs.} In this case, the direction of the $\gb$ is aligned with
the direction of {\bf B}\footnote{This is strictly necessary because \citet{obr07} has shown that the gradient increasing direction is parallel to the limb, so quasi-perpendicular injection efficiency implies {\bf B} perpendicular to shock normal at the limbs and therefore {\bf B} parallel to $\gb$,}, so the angle $\phigb$ is the same as $\phib$,
and we did not considered it any further. Another consequence of this is
that $R_{max}$ is always 1 (in disagreement with observations), so this parameter cannot give any diagnostics.
The results are shown in Fig. \ref{resqper}, and it seems that a good
fit can be found for $\phib \sim 70^\circ$ (a value in agreement with the
azimuthal profile analysis of \citealt{pdc09}), even if the model points in
the $\theta_D-C$ diagram seems to be more distant to the observed points
than any scatter plot in Fig. \ref{resqpar}, thus indicating a
better fit in the quasi-parallel case then in the quasi-perpendicular one.

\section{Discussion}

In this work, we introduced a new method of comparison between models and observations of radio SNRs shells that can provide useful information on the 3D structure of the remnant and on the unperturbed ambient magnetic field in which the remnant expands.
The method is based on the calculation of 4 morphological parameters.
In the case of SN1006, we found that the method gives a good fit if the injection efficiency is quasi parallel and the ambient magnetic field is characterized by a gradient of its strength. In addition, the method allowed us to constrain the viewing geometry angles of the magnetic field ($\phi_{\bf
B}=38^\circ\pm 4^\circ$ and $\phi_{\nabla |{\bf B}|}=15^\circ\pm 15^\circ$). 
{ \citet{svr10} showed that the quasi-perpendicular model is a better fit to the $R_{\pi/3}$ value (the ratio between the inner region and the limb region flux introduced by \citealt{rbd04}) of SN1006 in the radio band, but they did not properly considered aspect angles $\ne 90^\circ$ and the non-uniform magnetic field. We verified that our model gives  $R_{\pi/3}=0.16$ for $\phi_{\bf B}=90^\circ$, but $R_{\pi/3}=1.00$ for $\phi_{\bf B}=38^\circ$, in agreement with the value measured in the new radio map of Fig. \ref{radiomap} ($R_{\pi/3}^{obs}=1.01$)}.

\begin{figure}
  \centerline{\psfig{file=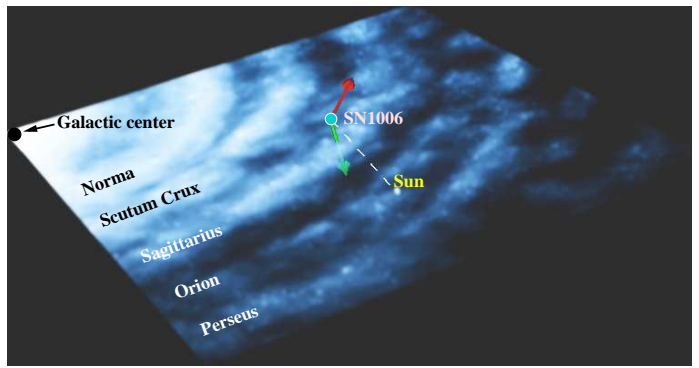,width=9.0cm}}
  \caption{Artist view of the magnetic field at the location of SN1006 in our Galaxy. The red and green arrows mark the directions of the field and its gradient as derived by the best-fit model for the synchrotron radio emission of SN1006, assuming a quasi-parallel scenario, derived in this work. The figure is in scale.}

  \label{magf3d}
\end{figure}

The symmetry axis of SN1006 is perpendicular to the galactic plane\footnote{This makes SN1006 rather peculiar, since the majority of bilateral SNRs have their symmetry axis parallel to the galactic plane (\citealt{gae98}).}, its distance is 2.2 kpc (\citealt{wgl03}) and its galactic latitude is $14.6^\circ$. Combining all this pieces of information, we can plot the direction of the magnetic field and its gradient in a 3D representation of the Galactic disk. This is shown in Fig. \ref{magf3d}.

Quite remarkably, our best-fit values of $\phi_{\nabla |{\bf B}|}$ and $\phi_{\bf
B}$ imply that the direction of $\nabla |{\bf B}|$ resulting from the fit to the quasi-parallel model points down toward the Galactic plane and the direction of {\bf B} is aligned with the direction of the spiral arm near the remnant. This is indeed a very reasonable configuration for the magnetic field of the Galaxy, so it is tempting to conclude that we are sampling the large scale field of the Galaxy. However, we note that the best-fit value of $\nabla |{\bf B}|$ (i.e. a variation of a factor 1.4 over 10 pc, model GRAD2 in Table \ref{models}) seems to be high if compared with the values predicted by current models of galactic MF. We have used the model of \citet{srw08} to compute the expected variation of the azimuthal field in the Galaxy at the location of SN1006. In particular, we have used their {\em ASS+RING} model which seems to be one of the favored by rotation measures of pulsars in the plane. At the SN1006 distance of $\sim 550$ pc from the plane, the large-scale field is expected to vary of a factor of 1.4 on scales of more then 100 pc, much larger than the scale-length required by our lowest gradient model (GRAD1 in Table \ref{models}). This opens up the possibility that we are actually sampling the random magnetic field component. \citet{ms96} reports length scales for this component of the order of few pc, which is compatible with the variations we derive.

\section{Summary and conclusion}

The synchrotron radio emission of the archetypical bilateral supernova remnant SN1006 is compared against a MHD model for this remnant, including a gradient of the magnetic field strength, particle acceleration and its obliquity dependent characteristic, namely quasi-perpendicular, quasi-parallel and isotropic scenarios for the injection efficiency and a rough treatment of magnetic field amplification. In order to explore the parameters space, which is very large due to the necessity to include the viewing geometry in the model-data comparison, we developed a simplified procedure based on the computation of four morphological parameters. We have found a very good fit with a model assuming quasi-parallel injection efficiency and
a configuration of the ambient magnetic field characterized by a variation of its strength of abut a factor of 1.4 (1.2--2.0) over a scale of 10 pc, and by viewing geometry angles 
$\phi_{\bf
B}=38^\circ\pm 4^\circ$ (between the direction of {\bf B} and the line of sight), and $\phi_{\nabla |{\bf B}|}=15^\circ\pm 15^\circ$ (between the plane of the sky and the direction of the gradient of the magnetic field). A worse fit is obtained in the quasi-perpendicular scenario. The overall morphology of the observed radio emission at 1.4 GHz is correctly recovered by our best-fit model. {  Therefore, we conclude that the SN1006 limbs are polar caps, significantly tilted with respect to the plane of the sky.}  The projected direction of {\bf B} and $\nabla |{\bf B}|$ in the Galaxy are along the spiral arm and toward the plane respectively, which is in very good agreement with the expected direction of the large scale galactic {\bf B}. However, the implied gradient is too high to be associated to the large scale galactic {\bf B} at the location of the SN1006 remnant and more typical of the random magnetic field components. 

The application of our method to selected samples of bilateral supernova remnants may yield independent estimates of the geometry of the galactic field at several locations, which can be useful to understand the field topology in our Galaxy.

\begin{acknowledgements}

FB acknowledges many fruitful discussions and the friendly environment at the ISSI meeting on "Magnetic field in the universe", and in particular with Aris Noutsos.
The software used in this work was in part developed by the
DOE-supported ASC/Alliances Center for Astrophysical Thermonuclear Flashes
at the University of Chicago. The calculations were performed on the cluster at the SCAN (Sistema di Calcolo
per l'Astrofisica Numerica) facility of the INAF – Osservatorio Astronomico di
Palermo.

\end{acknowledgements}

\bibliographystyle{aa}
\bibliography{../../references}

\end{document}